\documentclass{aa}
\usepackage{psfig}
\usepackage{epsfig}
\usepackage{graphicx}

\def \ss {AX~J0851.9$-$4617.4~}
\def \gg {G 266.1$-$1.2}

\def \hcm {\hbox {\ifmmode $ atom cm$^{-2}\else atom cm$^{-2}$\fi}}

\def\approxgt{\mathrel{\hbox{\rlap{\lower.55ex \hbox {$\sim$}}
        \kern-.3em \raise.4ex \hbox{$>$}}}}
\def\approxlt{\mathrel{\hbox{\rlap{\lower.55ex \hbox {$\sim$}}
        \kern-.3em \raise.4ex \hbox{$<$}}}}

\begin{document}

\title{An H$\alpha$ nebula possibly associated with the central
X--ray source in the \gg\ supernova remnant}

\author{A. Pellizzoni, S. Mereghetti, A. De Luca}

\offprints{A. Pellizzoni (alberto@mi.iasf.cnr.it)}

\institute{
        Istituto di Astrofisica Spaziale e Fisica Cosmica --
        Sezione di Milano ``G. Occhialini" -- CNR \\
         via Bassini 15, I-20133 Milano, Italy}

\date{Received ; Accepted:  }

\authorrunning{A. Pellizzoni et al.}

\titlerunning{{H$\alpha$ nebula }}

\maketitle

\abstract{We report the discovery of a small H$\alpha$ nebula 
positionally coincident with the candidate neutron star \ss 
located at the center of the supernova remnant \gg .
The nebula has a roughly circular shape with a diameter of 
$\sim$6$''$ and a flux of $\sim$$10^{-2}$ photons cm$^{-2}$ 
s$^{-1}$ in the H$\alpha$ line.
Considering the uncertainties in the distance and energy 
output from the putative neutron star, we find that such a 
flux can be explained either in a bow-shock model or assuming 
that the nebular emission is due to photo-ionization and heating
of the ambient gas.
}


\section{Introduction}
\label{sect:intro}

The X--ray source \ss is a strong neutron star candidate, very
likely associated with the shell-like supernova remnant \gg\ 
(Aschenbach 1998). \ss was first seen with the \textit{ROSAT} 
satellite (Aschenbach 1998, Aschenbach et al. 1999) and 
subsequently studied with \textit{ASCA} (Slane et al. 2001) 
and \textit{BeppoSAX} (Mereghetti 2001). Its location,  very 
close to the geometrical center of \gg\  , suggested an 
association with this young remnant, but the situation was
complicated by the presence  of two early type stars (HD 
76060 and Wray 16-30) that might have been responsible for 
the observed X--rays from \ss, as well as by the presence of
other X--ray sources in the vicinity (Mereghetti 2001).

The picture was finally clarified thanks to the accurate
localization obtained with the \textit{Chandra} satellite
(Pavlov et al. 2001).
The absence of any optical counterparts at the 
\textit{Chandra} position, down to magnitudes B$\sim$22.5 and
R$\sim$21, implies a very high X--ray to optical flux ratio, 
consistent with an isolated neutron star.
The soft  spectrum of \ss, well described by a blackbody with
$kT$$_{\rm{BB}}\sim$0.4 keV, is similar to that of other compact 
X--ray sources found in supernova remnants, such as, e.g.,  
CasA (Mereghetti, Tiengo \& Israel 2002) and G 296.5+10.0 
(Pavlov et al. 2002).
For an assumed distance of 1 kpc the X--ray luminosity of \ss
is $\sim$10$^{32}$$-$10$^{33}$ erg s$^{-1}$.
No  pulsations have   been detected so far.

Here we present optical images of the central region of \gg\ 
showing the presence of a small H$\alpha$ nebula at the position of \ss.

\begin{figure*}[!ht]
\vskip 0.0truecm
\centerline{\psfig{figure=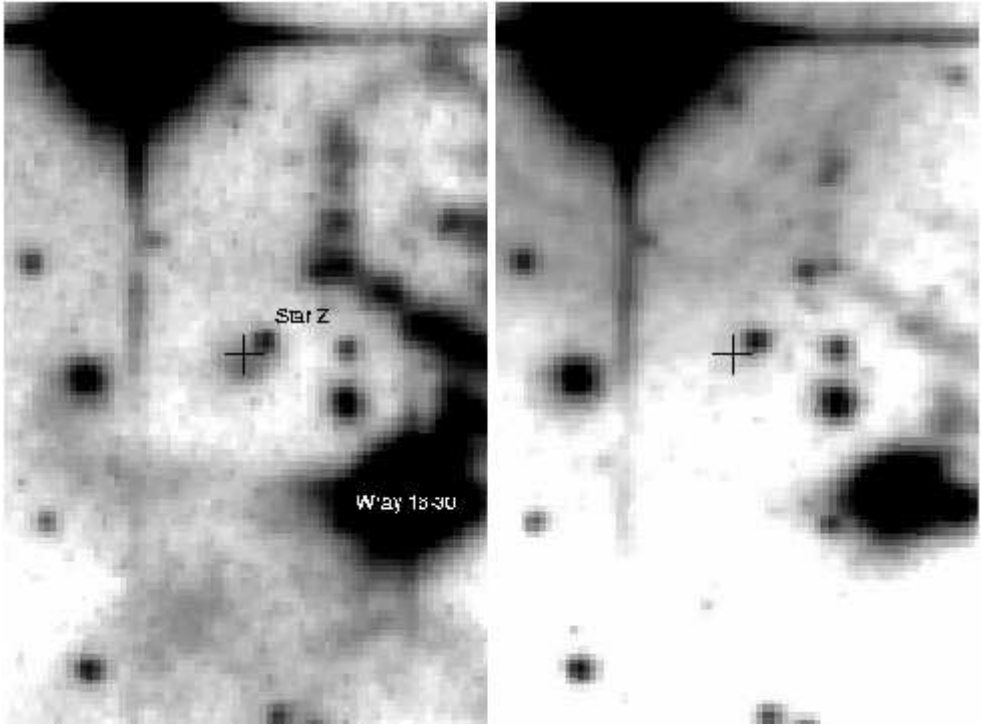,height=130mm,angle=-90}}
\vskip 0.0truecm
\caption{H$\alpha$ (left) and R band (right) images of the  central region
of \gg\, North is to the top, East to the left. The cross indicates the position of \ss
(RA(J2000) = 8$^{h}$ 52$^{m}$ 01$^{s}$.38, Dec(J2000)= $-$46$^{\circ}$
17$'$ 53$''$.34,  Pavlov et al. 2001).
To locate it, we performed an astrometry  based on the coordinates
of a number of stars from the GSC II catalog. The rms of our astrometry fit
is 0.07$''$; taking into account possible systematic errors in the GSC catalogue, we estimate
an overall error $<$1$''$.}
\label{fig:cfr}
\end{figure*}

\section{Observations}

We obtained images of the central part of \gg\ from the 
public archive of the European Southern Observatory (ESO).
The data were acquired with the Wide Field Imager (WFI) at the
2.2-m MPG telescope at La Silla.
They consist of 5 min long exposures in the R, B and H$\alpha$
filters taken on 2001 January 5.

To derive a photometric calibration we used the R and B band
magnitudes of several field stars as reported in the USNO A-2 
and GSC II catalogs (excluding objects with discrepant 
values).
We found that the star closest to the error circle of \ss 
(star Z of Pavlov et al. 2001) has magnitudes B$\sim$19 and
R$\sim$17, in agreement with the estimate of these authors.
The limiting magnitudes of the ESO WFI images are B$\sim$23 
and R$\sim$22.
No objects are visible in the \textit{Chandra} error region in these 
two bands.

The image in the  H$\alpha$ filter shows, besides a remarkable
nebula probably associated to the B[e] star Wray 16--30
(Th$\acute{e}$, de Winter \& P$\acute{e}$rez  1994), the 
possible presence of a slightly extended faint source at a 
position compatible with that of \ss.

The reality of such a source is confirmed by the deeper images
obtained with the UK Schmidt Telescope (UKST) at the 
Anglo-Australian Observatory as part of the H$\alpha$ 
survey of the Southern Galactic Plane and Magellanic Clouds 
(SHS, Parker \& Phillipps 1998).
This survey consists of long exposures (3 hr) in the H$\alpha$
filter and contemporaneous 15 min Short-Red (SR) broad-band 
exposures. These plates, digitised with the SuperCOSMOS 
facility (Hambly et al. 2001), provide an unprecedented 
combination of coverage
(6.5$^{\circ}\times$6.5$^{\circ}$), resolution ($\sim$1$''$) 
and sensitivity (5$\times10^{-17}$ erg cm$^{-2}$ arcsec$^{-2}$
s$^{-1}$).
A region \footnote{The data were kindly provided to us before
public release by Mike Read, Institute for Astronomy, 
Edinburgh, UK} of $\sim$0.8$'\times1.2'$ around the position
of \ss is shown in Fig. 1.
These data  clearly show a roughly circular nebula with
diameter $\theta$$\sim$6$''$ at the position of the X--ray source.
Such a nebula is only visible in the H$\alpha$ filter, while
it is not detected in the R band plates.
All the diffuse  H$\alpha$ emission is well highlighted in
Fig. 2, which shows the  ratio of the H$\alpha$ and R images.
The nebula, labeled with A, is coincident with the X--ray 
source and appears well separated from all the other diffuse 
features permeating this region.

We have derived the R$-$H$\alpha$ color of a number of stars 
and of  diffuse features also visible in the R image.  
There is some 
evidence that nebula A has a higher value of R$-$H$\alpha$  ($\approxgt$4) than
other diffuse features of comparable H$\alpha$ brightness.

Unfortunately, at this stage, the SHS on-line survey does not
provide a photometric calibration.
To estimate the flux of nebula A, we performed a rough 
calibration of the H$\alpha$ image as follows.
We first computed the H$\alpha$ fluxes corresponding to the R
magnitudes of a sample of GSC II stars present in our image, 
assuming blackbody spectra with the appropriate stellar 
temperatures.
We then derived the relation between the instrumental 
magnitudes and the H$\alpha$ fluxes.
This relation was found to have a small dispersion and allowed
us to transform the instrumental H$\alpha$ magnitude 
$-$11.8$\pm$0.2 of nebula A to a flux of $\sim$(1$\pm$0.2)
$\times10^{-2}$ photons cm$^{-2}$ s$^{-1}$ in the H$\alpha$ 
line.

\begin{figure}[!ht]
\vskip -0.5truecm
\hspace{1.0truecm}
\centerline{\psfig{figure=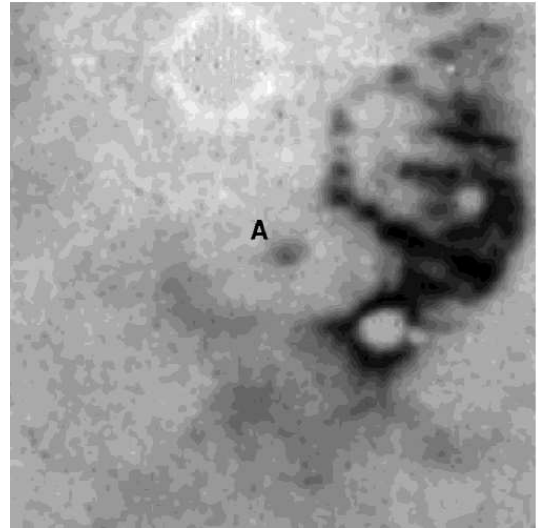,height=70mm,angle=0}}
\vskip 0.0truecm
\caption{ Ratio of the  H$\alpha$  and R band  images.
}
\label{fig:ratio}
\end{figure}

\section{Discussion}

``Balmer line-dominated'' nebulae have been detected close to 
four radio pulsars and to the radio-quiet neutron star 
RX J1856.5$-$3754 (see, e.g., Chatterjee \& Cordes 2002, and 
references therein).
Such nebulae, typically characterized by an arc-shaped or by a
``cometary'' morphology, can be explained as bow-shocks in the
ambient gas produced by the hypersonic motion of a neutron
star with a strong relativistic wind.
In this model, the expected H$\alpha$ luminosity is 
proportional to the neutron star velocity $v$, the rotational
energy loss $\dot{E}_{\rm{rot}}$, and the fraction $X$ of neutral 
hydrogen in the interstellar medium:

$$ L_{\rm{H\alpha}}~~ \propto ~~v ~\dot{E}_{\rm{rot}}~~X~~~~~~~   (1) $$

We have plotted in Fig. 3 the luminosity of the known 
H$\alpha$ nebulae associated to neutron stars versus the 
quantity $v\dot{E}_{\rm{rot}}$.
The sizes of the  rectangles reflect the possible range
of velocities allowed by the measurements and/or limits on 
proper motion, distance, and angle $i$ between the pulsar 
velocity and the line of sight available in the literature 
(see Table 1 for details).
After accounting for different plausible values of $X$ in the
range 0.01--1, these data are well explained by relation (1),
with the normalization factor derived by Cordes, Romani \& 
Lundgren (1993) (dashed lines).
The only exception is RX J1856.5$-$3754 for which the period is
unknown and therefore the $\dot{E}_{\rm{rot}}$ estimate is 
uncertain.

We have also indicated in Fig. 3 the luminosity of nebula A,
for assumed distances in the 0.5--2 kpc range (solid lines). 
It can be seen that an explanation in terms of the bow-shock 
scenario is, at least energetically, possible.
In this hypothesis, we can derive limits on $v$ and 
$\dot{E}_{\rm{rot}}$ as a function of the assumed values of distance and $X$.
Considering the small displacement from the SNR center ($\approxlt$10$'$) and the
lack of evidence for a high value of $\dot{E}_{\rm{rot}}$ 
(e.g. a bright X--ray or radio synchrotron nebula),
we are led to the conclusion that a relatively 
small distance is favored, unless the neutron star velocity is
close to the line of sight direction 
(see Pellizzoni et al. 2002 for more details).

\begin{figure}[!ht]
\vskip -0.5truecm
\centerline{\psfig{figure=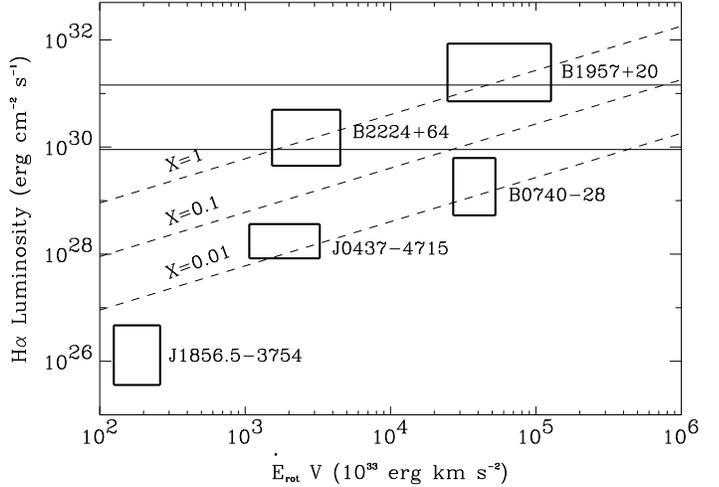,height=70mm,angle=0}}
\vskip 0.0truecm
\caption{H$\alpha$ luminosity versus $v\dot{E}_{\rm{rot}}$ for the
5 known H$\alpha$ nebulae associated to neutron stars.}
\label{fig:ratio}
\end{figure}

\begin{table*}
\begin{center}
\caption[]{References. - (1) Bell et al. (1995); (2) van 
Straten et al. (2001); (3) Jones, Stappers \& Gaensler (2002);
(4) Kulkarni \& Hester (1988); (5) Cordes, Romani \& Lundgren
(1993); (6) Chatterjee \& Cordes; (7) Kaplan, van Kerkwijk \&
Anderson (2002); (8) van Kerkwijk \& Kulkarni (2001).}
\begin{tabular}{cccccccc}
\hline
\noalign {\smallskip}
Name & H$\alpha$ Flux & Log $\dot{E}_{\rm{rot}}$ & $D$ & $v$$_{\rm{min}}$ & $v$$_{\rm{max}}$ & ref \\
 & ph/cm$^{2}$ s & erg/s  & kpc & km/s  & km/s &   \\
\hline
\noalign {\smallskip}
J0437$-$4715 & 2.5$\times$10$^{-3}$  & 34.07  & 0.139$\pm$0.003 & 90 & 280 & 1,2 \\
B0740$-$28 & 5$\times$10$^{-5}$  & 35.16  & 1.9$\pm$0.5 & 190 & 360 & 3 \\
B1957+20 & 1.09$\times$10$^{-2}$  & 35.20  & 1.5$\pm$0.4 & 160 & 800 & 4 \\
B2224+65 & 1.1$\times$10$^{-3}$  & 33.08  & 2.0$\pm$0.5 & 1300 & 3700 & 5,6 \\
J1856.5$-$3754 & 2$\times$10$^{-5}$  & 32.9$^{a}$  & 0.14$\pm$0.04 & 160 & 330 & 7,8 \\

\noalign {\smallskip}
\hline
\label{tab:spec}
\end{tabular}
\end{center}
a) assumed value (no period is known for this object).
\end{table*}

In an alternative interpretation, the nebular H$\alpha$ 
emission might be produced by photo-ionisation and heating of
the ambient gas by the extreme-ultraviolet radiation of the
neutron star. Such photo-ionisation nebulae were predicted to
exist around rapidly moving, hot neutron stars by Blaes et al.
(1995). Van Kerkwiijk \& Kulkarni (2001) proposed that this 
model could be responsible for the nebula associated to the 
nearby neutron star RXJ1856.5-3754.
Following these authors, considering a pure hydrogen gas 
nebula and a ionised fraction $X$$\sim$50\% for the emitting 
region, we find  that the blackbody emission with $kT$$\sim$0.4 
keV from \ss\ leads to a gas temperature $\sim10^{5}$ K due to
heating by photo-ionisation.
This implies plasma emission dominated by collisional 
excitation, with an expected H$\alpha$ photon rate, integrated
over the whole emitting region, given by:

$$ F_{\rm{H{\alpha}}}~~ \propto \frac{V~X~(1-X)~n_{\rm{H}}^2~q}{4\pi~D^{2}}~~~~ (2) $$

\noindent
where $V$$\sim$($\theta$$D$)$^{3}$ is the emitting volume and $q$ is the rate of
collisional excitations (Osterbrock 1989; Anderson et al. 
2000; van Kerkwijk \& Kulkarni 2001).
For $X$=0.5 and $q$=3.7$\times10^{-10}$ cm$^{3}$ s$^{-1}$,
appropriate for $T$=10$^{5}$ K, we find that the density 
required to match the observed photon rate of $\sim$10$^{-2}$
ph cm$^{-2}$ s$^{-1}$ is $n$$_{\rm{H}}\sim$5 $D$$_{\rm{kpc}}^{-0.5}$  
cm$^{-3}$.
Therefore, reasonable density values ($<$5 cm$^{-3}$) can 
only be obtained  for $D$$>$1 kpc.

\section{Conclusions}

We have discovered a faint ($\sim$10$^{-2}$ ph cm$^{-2}$ 
s$^{-1}$) H$\alpha$ nebula positionally coincident with the 
point-like X--ray source \ss , which is thought to be the 
compact remnant associated to SNR G~266.1$-$1.2.
Although this region
contains several diffuse H$\alpha$ features, the positional 
coincidence and the possible evidence for a peculiar color, 
compatible with a pure Balmer emission line spectrum, suggest
a relation between the H$\alpha$ nebula A and the putative 
neutron star.
In fact, although the lack of a  period and spin-down 
measurement for \ss make the energetics somewhat uncertain,
we have shown that the luminosity of the nebula is compatible
with the predictions of the two models which have been invoked
to explain a few H$\alpha$ nebulae associated to different
kinds of neutron stars.

Different distances are favored by the two scenarios. 
The model based on photo-ionization suggests  a distance greater
than $\sim$1 kpc, consistent with that inferred from X--ray absorption 
in the \gg\ SNR (Mereghetti \& Pellizzoni 2001).
In the case of a 
bow-shock nebula, the  most likely distance is 
smaller than $\sim$0.5 kpc, due to the presumably small value
of $\dot{E}_{\rm{rot}}$  and the lack of evidence for a high 
transverse velocity for the neutron star. 
Of course we cannot exclude a greater distance in the bow-shock scenario 
if the neutron star  has a high velocity close to the direction of the 
line of sight. 
Although this might be consistent with the circular symmetry
of the  nebula, we note that the chance probability for, e.g., 
$i<10^{\circ}$ is only $\sim$7\%.

More detailed investigations are required to confirm the 
proposed association between nebula A and \ss\, and 
eventually to discriminate between the two possible 
mechanisms.
In particular, deeper imaging with high resolution can provide
information on the shape of the nebula, while high-resolution
spectroscopy is needed to measure the width of the H$\alpha$
line.
In the bow-shock model one should see a major fraction of the
emission with velocity widths comparable to the shock 
velocity, larger than that of the narrow lines expected in the
photo-ionisation model with thermal velocities 
$\approxlt40$ km s$^{-1}$ (Raymond 1991).

\begin{acknowledgements}
We used public data obtained from the ESO/ST-ECF Science 
Archive Facility.
SuperCOSMOS facility and AAO/UKST survey staff (in particular,
Mike Read and Nigel Hambly) kindly provided us prompt 
technical assitance to analyse data.
\end{acknowledgements}

\end{document}